\documentstyle[epsfig]{mn}
\def\be{\begin{equation}}
\def\ee{\end{equation}}
\def\msun{\mbox{M$_\odot$}}

\title[Chemical Evolution Models of Local dSph Galaxies]
{Chemical Evolution Models of Local dSph Galaxies}
\author[L. Carigi, X. Hernandez and G. Gilmore]
{Leticia Carigi $^{1}$, Xavier Hernandez $^{1}$, and Gerard Gilmore$^{2}$\\
$^1$ Instituto de Astronom\'\i a, Universidad Nacional Aut\'onoma de
M\'exico, A.P. 70-264, 04510 M\'exico, D.F. \\
$^2$ Institute of Astronomy, Madingley Road, Cambridge CB3 0HA, England\\
}

\date{\today}

\begin{document}
\maketitle

\begin{abstract}

We calculate chemical evolution models for 4 dwarf spheroidal satellites of the Milky Way
(Carina, Ursa Minor, Leo I and Leo II) for which reliable non-parametric star formation histories 
have been derived. In this way, 
the independently obtained star formation histories are used to constrain the
evolution of the systems we are treating. This allows us to obtain robust inferences on the
history of such crucial parameters of galactic evolution as gas infall, gas outflows
and global metallicities for these systems. We can then trace the metallicity and abundance ratios
of the stars formed, the gas present at any time within the systems and the details of gas ejection, 
of relevance to enrichment of the galaxies environment. We find that galaxies showing one single
burst of
star formation (Ursa Minor and Leo II) require a dark halo slightly larger that the current estimates 
for their tidal radii, or
the presence of a metal rich selective wind, which might carry away much of the energy output of their
supernovae before this might have interacted and heated the gas content, for the gas to be retained 
until the
observed stellar population have formed. Systems showing extended star formation histories (Carina and 
Leo I) however, are consistent with the idea that their tidally limited dark haloes provide the 
necessary gravitational potential wells to retain their gas. The complex time structure of the star 
formation in these systems, remains difficult to understand. Observations of detailed abundance ratios
for Ursa Minor, strongly suggest that the star formation history of this galaxy might in
fact resemble the complex picture presented by Carina or Leo I, but localized at a very early epoch.

\end{abstract}


\section{Introduction}

Our understanding of the dwarf spheroidal companions of the Milky Way has advanced significantly
over the last few years. Studies of the internal dynamics of stars measuring velocity dispersions,
have revealed the presence of varying but always significant amounts of dark matter 
(e.g. Mateo et al. 1993, Tamura et al. 2001), strengthening
the classification of these systems as galaxies. The structure of the dark haloes associated with 
these galaxies appears to be well represented by a constant density region, over the extent across which
measurements exist. High quality imaging of the stellar populations
has permitted the construction of HR diagrams for many of these galaxies, which in turn has stimulated
the development of new statistical analysis techniques aimed at recovering the star formation histories
of the imaged systems e.g. Aparicio \& Gallart (1995), Tolstoy \& Saha (1996), Mighell (1997),
Hernandez et al. (1999). These last have yielded valuable constraints on the temporal 
structure of the star formation histories in the local dSph population. Results have shown there to be
a wide range of different star formation histories here. Some systems being consistent with a single
burst of activity in the remote past, but many showing clear signs of extended or repeated periods 
of star formation activity.

The dominance of dwarf galaxies in cluster luminosity functions indicates that the collective
contribution of these systems to the chemical evolution of clusters is quite probably significant.
Detailed studies of the role of small galactic systems in the context of the enrichment of the 
inter-galactic medium in clusters have confirmed the above, generally treating dwarf and dSph galaxies 
as small scale analogues of elliptical galaxies, typically characterized by a single burst of star 
formation in the remote past, and nothing since e.g. Gibson \& Matteucci (1997). 
Recently, the prevalence of higher than primordial metal abundances in the high redshift $Ly\alpha$
systems was taken by Madau, Ferrara \& Rees (2001) as an unequivocal indication of substantial
metal pollution by galactic systems of masses in the dSph range at high redshift. The 
details of the star formation physics and efficiency in these small systems remains as a free
parameter which influences the details of their calculations significantly, again approximated
by a single burst of activity.
However, given the
complex star formation histories local dSph's present, it is worthwhile to analyze more closely what
a careful tracing of the chemistry in these systems, fixed so as to reproduce their observed star 
formation histories within their observed dark haloes, predicts in terms of metallicities, abundance 
ratios and total amounts of ejected gas. The smallest galactic structures also play a determinant part 
in models of reionization of the universe, where again, these systems are usually treated only in a 
very generic way (e.g. Ciardi, Ferrara \& Abel 2000). 
The local dSphs treated in this work offer a unique 
opportunity to study the details of galactic evolution at the smallest scales, and hence calibrate 
the details of their contribution to a variety of astrophysical problems.

The observation of complex stellar populations in the dSph satellites of the Milky Way, in the total
absence of gas (e.g. Bowen et al. 1997), implies that the gas which was at some point present
in these systems to fuel star formation processes, has now been totally lost. Given their 
dynamical masses, theoretical models of gas heating due to supernovae (henceforth SN) and stellar 
winds in these systems predict the efficient formation of galactic
winds after only a few hundred Myrs of star forming activity, with the resulting loss of all gas
(e.g. Dekel \& Silk 1986, Ferrara \& Tolstoy 2000).
As already pointed out by Gallagher \& Wyse (1994), the low escape velocities of dSph's implies that
the evolution of gaseous and stellar metallicities will be determined mostly by the inflows and 
outflows of gas, rather than by the details of the star formation history, as is the case in larger
systems, where the use of closed box models is generally accepted as a good first approximation.
Note however, the recent detection of some HI in Sculptor by Carignan et al. (1998), a galaxy
with some indication of a young stellar component superimposed on a dominant old population. 
The distance to Sculptor does not allow the detailed study of its stellar population, so that
the inferences of its SFH are only tentative, and doubts remain concerning the actual physical
association of the observed gas to the galaxy. This again stresses the poor knowledge of the
interstellar medium-star formation rate (henceforth ISM and SFR, respectively) 
connection in these systems, a consequence of the absence of a detailed physics for 
star formation.

The presence of stellar populations showing age spreads of several Gyrs, remains the greatest puzzle
surrounding these small galaxies. Re-accretion of gas has been proposed as the only plausible
explanation for the extended star formation histories observed, although the details and causes
of any repeated periods of accretion also remain a mystery. In this work we aim at restricting the
parameters describing repeated periods of gas accretion in some of the local dSph galaxies, as well
as obtaining indications as to the evolution of the metallicity in them.

We construct a chemical model for the evolution of the stellar and gaseous constituents of
a small galactic system, paying attention to the enrichment processes, and the heating and dynamics of
the gas. The dynamics being largely determined by the dark halo of the systems,
which we fix from independent inferences. The determinant input of the model is now the accretion
history of fresh gas, with the metallicity of the stars and gas, together with the appearance of 
a galactic wind being the output of the model. In absence of a detailed formation and
evolution scenario, this gas accretion history must be treated through free parameters. 
Fortunately for 4 of the local dSph's, direct and objective determinations of
their star formation histories exist. Hernandez, Gilmore \& Valls-Gabaud (2000), henceforth HGC, 
use a rigorous maximum likelihood
statistical method to derive the SFR(t) for 4 dSph galaxies, Carina, Ursa
Minor, Leo I and Leo II, in a totally non parametric approach. 

In a complementary approach, Carraro et al. (2001) model dSph systems from first principles, and show 
that for systems having
a fixed mass, in the dSph range, variations in the total dynamical density can result in a range
of different star formation histories, spanning that presented by the satellites of the Milky
Way. Again, the star formation process was introduced through dimensional
arguments and free parameters, in the absence of detailed knowledge of star formation physics, and the
galaxies are treated as isolated systems. This last assumption probably does not hold for local dSph's
where the presence of the MW exerts a possibly dominant effect on the evolution.
The influence of the tidal field of the Galaxy is well established in several of these systems, 
Martinez-Delgado et al. (2001) find evidence of tidal tails in Carina, and Majewski et al. (2000) 
determine the presence of similar streams of drawn out material in Ursa Minor. Further,
Hirashita, Kamaya \& Mineshige (1997) point to the dynamical frailty of the gas in dSph systems
in relation to the gas loss associated to the hot outflows produced by the star formation processes
in the disk of the Milky Way. Scanapieco, Ferrara \& Broadhurst (2000) confirm the above through
detailed dynamical modeling of the photo evaporation and ram striping of gas from dSph's due to
galactic winds and fountains. All this only stresses the fact that these deceivingly simple
galactic systems are subject to complex processes which make it difficult to construct physical
models to describe their evolution.

In our present study we take the SFR's 
derived by HGC (henceforth $SFR_{HGV}$) as external constraints on our chemical evolution 
models, and hence obtain interesting restrictions on the
time structure and magnitude of the gas accretion history of Carina, Ursa Minor, Leo I and Leo II.
In a sense, for these galaxies we know part of the answer in advance, and can hence calculate
the energy input produced by the inferred star formation history, and restrict the possible
gas accretion and outflow scenarios, thus obtaining valuable information on the nature of the
ISM-SFR connection.
The time evolution of the metallicities is then a prediction of the model, which we can compare with 
observed values.

The models we obtain show a variety of possibilities for the physical evolution of these systems,
depending on which parameters one varies to ensure gas is retained until the luminous galaxy is formed.
Galaxies showing repeated periods of star formation, such as Carina and Leo I in our sample, can only be
explained with the inclusion of a re-accretion of fresh gas. We obtain predictions on the total masses,
metallicities and abundance ratios of the ejected material, as a result of having carefully traced the
physics of the gas content, the different SN yields, and the final results of the outflows. A simple 
physical criterion is also proposed as relevant to discriminating dSph galaxies subject to extended and 
repeated star formation, from those susceptible only to a single burst of activity.

The plan of our paper is as follows: Section 2 presents the details of the enrichment and gas dynamics 
model, with the results once the inferred SFR(t)'s have been introduced as constraints, 
presented in Section 3. Finally, a discussion of our results is given in Section 4 and Section 5 
states our conclusions.

\section{Theoretical Model}
\label{sec:model}

As mentioned in the introduction, with the possible exception of Sculptor, all attempts at 
detecting the presence of gas in dSph's have
yielded only null results. It is therefore reasonable to assume that the heating and
dynamical effects of star formation have powered winds which resulted in the loss of gas
in these systems. We shall assume that only SNae type I and II are responsible for these heating 
and dynamical processes, and calculate the appearance of galactic winds in dSph's accordingly.
The criterion for the establishment of a wind in essence derives from a comparison of the
thermal energy of the gas and its gravitational binding energy. It is hence the structure of the
dark matter halo which fixes the boundary conditions on the problem. In the following sub-section
we describe the details of the dark matter haloes used, and the criterion used to identify the
formation of a wind.

\subsection{Gravitational Assumptions}

We have assumed that a dwarf spheroidal galaxy is a system made initially
of a non-baryonic dark matter  halo
and a baryonic gas spheroid.
Direct studies of rotation curves in dwarf galaxies have shown the density
profiles of these systems to be well described by a constant density core,
followed by an isothermal region out to the limit of the observations 
(e.g. Burkert 1995). Observations in low surface brightness galaxies
have shown the same results (e.g. de Blok \& McGaugh 1997), and indeed the pattern appears to extend
to the dark components of clusters of galaxies imaged in X rays (e.g. Firmani et al. 2000).
It has also been shown that high surface brightness galaxies are also consistent
with this halo structure (Hernandez \& Gilmore 1998a), which it hence seems reasonable
to assume as universal. For the systems we are treating here, the details of the dark halo
beyond the core radius are largely unimportant, as their presence in the halo of the
much larger Milky Way implies the existence of a tidal radius for these galaxies, beyond which
the tidal field of our Galaxy tears off material. It can be shown that these tidal radii
are in fact very similar to the core radii of the visible dSph's. We therefore take a dark matter (DM)
distribution represented by a constant density out to the tidal radii of each dSph,
followed by an exponential cut-off starting at core radius (Hernandez \& Gilmore 1998b).

\be
\rho_{DM}(r) = \cases{ \rho_{D0} & if $ r < R_{core}$ , \cr
                    \rho_{D0} exp(1 - r/R_{core}) & if $ r > R_{core}$ , \cr}
\ee

\noindent
where $R_{core} = f_{DM} R_{t}$,  $R_t$ is the tidal radius and $f_{DM}$ is 
an input parameter in each model, varying which with respect to unity we can asses the
dependence of our results on the details of the dark halo within which the galaxies are embedded.
Indeed, Odenkirchen et al. (2001) perform a new survey of Draco using the Sloan Digital Sky Survey,
and conclude that the true core radius of the light distribution is 40\% larger than previous 
estimates showed, with no evidence of any tidal features. This shows that values of tidal radii
found presently in the literature could be lower limits in other cases as well. Also,
present day values for this parameter should be considered lower limits for the corresponding time 
averaged quantities, since the tidal field of the MW could well have reduced the dark haloes
of dSph over time.
The details of the density cut beyond $R_{core}$, indeed, the presence of any dark matter beyond 
$R_{core}$ only marginally affect our results.

The initial gas distribution can be deduced from the present day light distribution,
by assuming that the stars map the density distribution of the gas from which they formed.
From the observed exponential surface density light profile seen in dSph's
we infer for the gas distribution:
\begin{equation}
\rho_{g}(r) = \cases{ \rho_{g0} & if $r \leq R_g$ , \cr
                    \rho_{g0} exp(-(r-R_{g}/R_e) & if $r > R_g$ , \cr}
\end{equation}
\noindent
where $R_g$ is the observed core radius of the luminous galaxies and $R_e$ is the exponential 
scale length of the surface brightness distribution. In all that follows we will be assuming
that the form of both gas and stellar profiles is the same, with the respective normalizations
given by the relative amounts of both components.
Notice that as the potential energy is determined overwhelmingly by the dark halo,
the onset of the galactic wind is sensitive to the total gas present, but depends very 
little on the details of the gas distribution.

All structural parameters ($R_{t}$, $R_g$, $R_e$ ) are taken from Mateo (1998) and
are assumed fixed during the evolution of the systems we study.

Since current studies show dSph galaxies contain only stars, the present-day dynamical mass
is formed by DM  and stellar mass. The stellar mass is computed from the total luminosities of the
galaxies and the respective $SFR_{HGV}$, assuming a constant IMF.
Mateo (1998) gives the dynamical mass value
at $R_t$, so  when $R_{core} > R_t$ the dynamical mass is scaled by $(R_{core}/ R_t)^3$
since we have assumed constant density for the core regions.
 
The initial  baryonic (gas+stars) to non-baryonic mass ratio is indicated by $M_{lum}/M_{DM}$,
which is an input parameter. By default, $M_{lum}/M_{DM}=0.09$, that is the maximum baryonic content
of late type galaxies obtained by Hernandez \& Gilmore (1998a). Notice that this value refers only
to material within the core region, large amounts of dark matter would have been contained in the
initial halo, much of which has been shaven off by the Galactic tidal field, not so the baryons, 
which cooled and contracted at very early times.

We define $E_{GRAV}$ as the potential energy of the gas, in this case, the energy necessary 
to carry all the gas in the system out to a radius $= R_{core}$ (Martinelli, Matteucci,\& 
Colafrancesco 1998), at which point it will be stripped by the tidal field of our Galaxy as:

\begin{equation}
E_{GRAV}= \int_0^{R_{core}} \left( \int_r^\infty F(\xi) d\xi \right) dm_{g}(r) 
\end{equation}

where $F(\xi)$ is the force between $dm_{g}(r)$ and the dynamical mass for $r < \xi$,

$$F(\xi)=G \frac{(M_{s}(\xi)+M_{g}(\xi) + M_{DM}(\xi))}{\xi^2}$$
 and
$$ dm_{g}(r)= 4 \pi r^2 \rho_{g}(r) dr$$

From the above, we see that the exponential cut off in the dark matter density is not considered
in the dynamical determination of the problem, and has only a small relevance in determining the total
amount of baryonic matter, through the $M_{lum}/M_{DM}$ ratio. We assumed that 
$M_{DM}$ is not affected by internal or external 
galactic processes, as it is dynamically dominant, and is subject only to gravitational interactions.
We further assume that the shape of $M_{g}(r)$ and $M_{s}(r)$
is conserved during the whole evolution. The gravitational potential of the gas
evolves according to the evolution of $M_{g}(r)$ and $M_{s}(r)$. It will be the total gravitational
potential energy of the gas that will establish its stability, when compared to the thermal energy
of this component. In this way, the determinant factor is the escape velocity, which is an integral
property of the dark halo ({\it viz} Eq. 3). This makes our results rather insensitive to the details
of the dark matter profile taken. We note the recent results of Kleyna et al. (2001) who use a 
new maximum likelihood analysis together with recent measurements of stellar kinematics in Draco to
model the dark matter halo of this galaxy. These authors find again large amounts of dark matter, 
and a density profile well fitted by an isothermal $\rho(R) \propto R^{-2}$ almost into the very centre. 
The resulting escape velocities being much the same as obtained from our Eq. 3.

To compute the thermal energy of the gas ($E_{THER}$), it is necessary to know the number of SN
events occurring in our galaxies. We will use a chemical evolution code, which
computes the SN rates, abundances,  gas and stellar masses, as a function of
gas accretion and star formation rates, as a function of time.
  
\subsection{Chemical Assumptions}

We have considered a one-zone chemical evolution model under the following assumptions:

1) The baryonic component of dSph galaxies is formed by an infall, A, of primordial material
	($X_o=0.76$, $Y_o=0.24$). 
a) The initial infall rate decreases exponentially with time.
$$\dot{A}=A_0 e^{-t/\tau}$$
$A_0$ is determined by requiring that the present day total luminosity of the models matches observed
values. $\tau$ is another input parameter and it is determined by the begining of the first
star formation episode. In cases with more than one episode, a secondary infall begining 1 Gyr before
the second burst is assumed.
b) Since Carina and Leo I have two main star formation episodes, 
a constant secondary infall is considered.
The duration and intensity of this infall are free parameters and are chosen such that
the second galactic wind occurs after the maximum of the second SF burst.

2) When $E_{THER}=E_{GRAV}$ a wind develops and efficiently clears the galaxy of gas.
Two types of outflows are assumed:
a) A well-mixed one in which galaxies eject all gas present to the intergalactic medium via a
galactic wind, for galaxies characterized by a single episode of star formation (burst galaxies) 
or via two galactic winds, for those showing also a secondary episode (complex galaxies), respectively.
b) A metal-rich one, where a fraction $\gamma$ of the mass of a SN (II, Ib, or Ia)
is ejected to the intergalactic medium without mixing with the ambient interstellar gas

3) The star formation rate, $\Psi$, is proportional to that inferred by 
HGV, 
$$\int_{0}^{15} SFR_{MODEL}(t) dt= \nu \int_{0}^{15} SFR_{HGV}(t) dt .$$ 
The proportional constant
is denoted by $\nu$. Values of $\nu$ distinct from unity were explored to test the dependence of
our results with respect to uncertainties in the total luminosities and IMF of the objects studied,
which determine the normalization on $SFR_{HGV}$.

4) The initial mass function, $\Phi$, is that of Kroupa, Tout \& Gilmore (1993),
between 0.08 \msun and 120 \msun. Shape and mass range are identical to those adopted by 
HGV.

5) We adopted metallicity dependent stellar properties in a code without any instantaneous 
recycling approximation, with $\tau_{s}(m)$  being the lifetime of stars, as a function of their mass
and metallicity. Specifically, yields and remnant masses are taken from
Maeder (1992)  for $9 \leq m/\msun \leq 120$, and from Renzini \& Voli (1981) for
$1 \leq m/\msun \leq 8$, with stellar winds contributing to the yields.

6) We assumed three types of SN: Ia, Ib, and II.
The supernovae rates are computed as described in Carigi (1994).
In this work
the fraction of binary systems predecessors of  SNIa is assumed as $A_{bin}=0.07$
and SNIb produced by binary systems are not considered (Carigi 2000).

We define $E_{THER}$ as the thermal energy of the gas from supernovae,
\begin{equation}
E_{THER}(t) = \int_0^t{\epsilon(t-T)RSN(T)dT}
\end{equation}
where $\epsilon(t)$ is the evolution of the thermal energy in the hot, dilute interior
of the supernovae remnants.

$$\epsilon(t)=\cases{0.72\epsilon_0 & if $t \le t_C$;\cr
                            0.22 \epsilon_0 (t/t_C)^{-0.62} &  if $t \ge t_C$ \cr} $$
where $\epsilon_0=10^{51}$ erg and $t_C$ is a cooling time scale taken from Cox (1972).
The above timescale was derived for solar metallicities, a more complete treatment of the cooling
processes in the expanding SN shell is given in e.g. Tantalo et al. (1998). In that work the cooling
timescale we used here is compared to more detailed metallicity dependent ones. These cooling
timescales basically give the time it takes for the energy of the supernova to be thermalized by
the surrounding inter stellar medium. The detailed timescales vary from about $2 \times 10^4$ years
for solar metallicities, to about  $2 \times 10^5$ years for metallicities in the range of what is 
found in dSph galaxies. It is clear that since the timescales for star formation are of the order
of Gyrs, whether each supernova completes its heating processes in $2 \times 10^4$ or  
$2 \times 10^5$ is irrelevant to the problem. We did in fact run a few models with the complicated
metallicity dependent scheme of the above authors and found it made no difference for our particular
problem. For simplicity we used the time scale from Cox (1972) throughout.

$RSN(T)$ is the supernovae rate, and includes SNIa, SNIb, and SNII.
SNIa originate from C deflagrations in a C-O white dwarf forming a binary system
with a red giant star, we compute type I supernovae rates following
Greggio y Renzini (1983):

$$
{\rm RSNIa}(t) = A_{bin} \int_{2.65}^{15} {\Phi(m_B)\over m_B} \int_{max\lbrace m_t(t), m_B-7.5 
\rbrace}^{0.5m_B}{f(m_2)}
$$
$$
{\times \Psi(t-\tau_{s}(m_2))dm_2}  dm_B, $$

where $m_t$ is the mass of the most massive star in the main sequence from the first stellar generation,
$m_2$ is the mass of secondary star, $m_B$ is the binary system mass and
$f(m_2)$ is the binary mass distribution function,
$$f(m_2) = 24 ({m_2\over m_B})^{2} $$

SNIb originate from W-R stars, and 
since the lower mass limit for W-R star progenitors ($lim_{WR}$) depends on the stellar metallicity 
(Maeder 1992) then, 

$$
RSNIb(t)= \int_{max \lbrace m_t(t),lim_{WR} \rbrace }^{120} {b(m,t)\Phi(m)\Psi(t-\tau_{s}(m))dm}
$$
where 
$$ b(m,t) =\cases{1&if $lim_{WR}(Z(t-\tau_{s}(m)))\le m/M_{\odot}  \le 120$;\cr
                            0&  for the rest \cr} $$

SNII arise from the core collapse of massive single stars between 7.5 \msun and $lim_{WR}$, therefore

$${\rm RSNII} (t) = (1 - A_{bin})\int_{max\lbrace m_t(t),7.5 \rbrace}^{15} {c(m,t)\Psi(t-\tau_{s}(m))\Phi(
m)dm}$$
$$
 + \int_{15}^{lim_{WR}} {c(m,t)\Psi(t-\tau_{s}(m))\Phi(m)dm}
$$
where,
$$ c(m,t) =\cases{1&if $7.5 \le m/M_\odot \le lim_{WR}(Z(t-\tau_{s}(m)))$;\cr
                            0&  for the rest \cr} $$

In forcing the resulting SFR(t) of a model to match the inferences of HGV some
details must be noted. The method used by these authors yields results which
are subject to a time varying error, which due to the decreasing separation of 
stellar isochrones with increasing age, compounded by the increasing observational
errors with increasing magnitude, ( and hence towards the turnoff of older populations)
increases significantly with age. Actually, the time structure is essentially lost for all
ages greater than 10 Gyr, corresponding to the first 5 Gyrs in the history of these systems,
in the axis shown in Figure 1. Also, given the total cessation of star forming activity after a strong
galactic wind which we are assuming, the time structure of $SFR_{HGV}$ was slightly modified
and re-normalized to give the observed present day total luminosities.

\section{Results}

As it can be seen from the large number of parameters needed to specify a particular model, 
our knowledge of the detailed structure and evolution of dSph's remains quite poor. We hence perform
an exploration of parameter space, looking for results which are robust with respect to the
details of the model, and seeking correlations between the model parameters, which 
reflect the underlying physics of these galaxies.

We have tuned the gas accretion parameters of the generic model
by requiring that given the SFR(t) inferred by HGV for each
galaxy, galactic winds fully clear the systems of gas only once the luminous galaxy has formed.

The results of HGV can be divided into two groups, with Leo II and Ursa Minor showing
a single episode of star forming activity. One expects that the DM halo of these galaxies 
manages to retain their gas only up to the onset of a galactic wind, which ends all SF activity.
A second group is formed by Carina and Leo I, which
show extended and repeated episodes of star forming activity. In these last two cases, 
the wind criterion can only be satisfied by the inclusion of at least a second episode of gas infall.   
Finally, for the case of the more nearby Carina and Ursa Minor galaxies, the results of HGV refer
strictly only to the portion of the galaxy sampled by the observations, local variations might exist 
which would make this different from the average over the whole galaxy.

\subsection{ Burst Galaxies}

\begin{table}
\caption{Model Parameters for Burst Galaxies}
\begin{center}\scriptsize
\begin{tabular}{lcccccc}
Model &  $f_{DM}$ & $\nu$ & $\gamma$ & $M_{lum}/M_{DM}$ & $\tau$ & 
$\Delta\gamma_Z$\\
\hline
\hline
&& Leo II &&&& \\
Leoii.dm  & 1.28 & 1.00 & 0.00 & 0.09 & 2.0 & [0.48, 0.89]\\
leoii.sfr & 1.00 & 0.24 & 0.00 & 0.09 & 2.0 & [0.00, 0.31]\\
leoii.wind& 1.00 & 1.00 & 0.82 & 0.09 & 2.0 & [0.07, 0.16]\\
leoii.bar & 1.00 & 1.00 & 0.00 & 0.28 & 2.0 & [0.20, 0.84]\\
\hline
&& UM &&&& \\
um.dm  & 1.32 & 1.00 & 0.00 & 0.09 & 0.001 & [0.77, 0.93]\\
um.sfr & 1.00 & 0.31 & 0.00 & 0.09 & 0.001 & [0.15, 0.76]\\
um.wind& 1.00 & 1.00 & 0.74 & 0.09 & 0.001 & [0.16, 0.23]\\
um.bar & 1.00 & 1.00 & 0.00 & 0.24 & 0.001 & [0.63, 0.89]\\
\hline
\hline
\end{tabular}
\end{center}
\begin{flushleft}
Columns 2-6 give the values the given parameters take when enforcing the
condition that no wind
develops until the peak of the inferred SFR is reached.
$f_{DM}$ gives the value by which the tidal radius was extended.
$\nu$ gives the factor by which the inferred star formation rate had to be
multiplied by.
$\gamma$ is the fraction of the energy input of the supernovae which was
assumed to be carried off (together with a corresponding fraction of metals)
by the wind before interacting with the ISM of the galaxies.
$M_{lum}/M_{DM}$ is the value used for the initial ratio of baryonic to
dark matter in the
models. In each of the models only one of the above 4 parameters was
changed with respect to its
zero order inferred/guessed value, any number of alternative models having
(almost) linear combinations
of the values showed above are also possible.
$\tau$ is the timescale of gas inflow accretion, basically fixed by the
onset of
star formation in the galaxies studied.
$\Delta\gamma_Z$ is the fraction of metals (only) which have to be assumed as lost from the system,
over what each model produces,
if we want the final average metallicities to match observed values, for each model.
\end{flushleft}
\end{table}

\begin{figure}
\epsfxsize=8.5cm 
\epsfbox{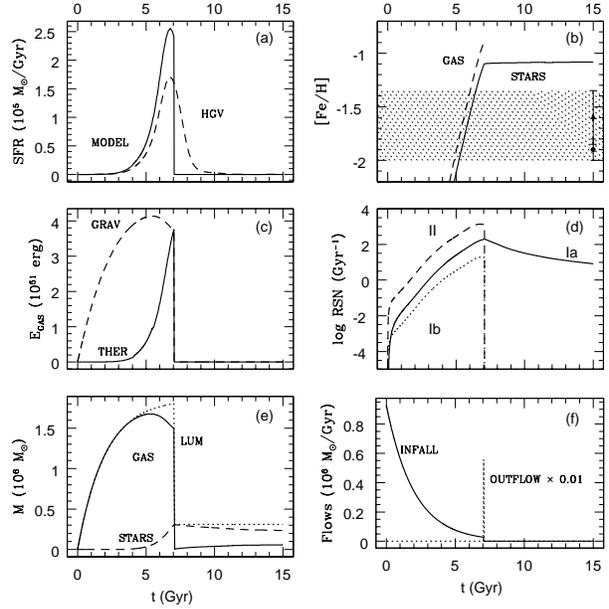} 
{\caption[]{
Predictions from leoii.dm model. 
(a) History of the star formation rate, according to HGV (dashed line)
and that assumed by the model (continuous line).
(b) [Fe/H] time evolution, gas abundance (dashed line) when SFR $\ne 0$ and mean abundance ratio
for stars between 0.5 and 2 \msun, solid curve. Data from Mateo (1998)
and Mighell \& Rich (1996) are shown by the hashed region, with the measured values and their
uncertainties arbitrarily placed at t=15 Gyr, circle and triangle, respectively. Solar abundances as
by Grevesse \& Sauval (1998).
(c) Evolution of the potential energy of the gas (dashed line) and thermal energy (continuous line).
(d) Theoretical supernova rates as a function of time. SNII (dashed line), SNIb (dotted line),
and SNIa (continuous line).
(e) Mass evolution. Gas mass (continuous line), stellar mass (dashed line), baryonic mass (dotted line) 
(f) Flows evolution. Accretion (continuous line), galactic wind (dotted line) 
}
\label{f1}}
\end{figure}

Table 1 gives the values of the input parameters of 4 models calculated for Leo II and Ursa Minor.
The first row shows a model where we have tuned the dark matter halo of the galaxy
so that no galactic wind develops until the maximum of the inferred SFH is reached, leoii.dm. The value of
$\tau$, the parameter which describes the accretion of gas, is fixed by the inferred onset of
the SF episode of this galaxy. Since $\gamma =0$, all the energy and metals of all the SNae are
mixed into the gaseous content of the galaxy. $f_{DM}$, the factor by which the 
tidal radius is multiplied to give the core radius of the dark matter halo in this case, 
must be fitted to 1.28. In this way, assuming the SFR(t) inferred by HGV, we find that we require a
dark halo slightly larger than the tidal radius of this galaxy, if we want to form the stars it formed
in the time it formed them. The details of the accretion formula are largely unimportant, 
any other infall prescription having the duration required by the SFR(t) inferences, would give 
basically the same result. Since a core radius 28\% larger than the tidal radius was required, 
we can conclude that not all the energy associated with SF can have participated in heating the ISM,
or that estimates of $R_{t}$ are off by 28\% for Leo II, not terribly unlikely.

To put this 28\% in perspective, the second row of the Leo II models has $\gamma=0$, $f_{DM} =1$,
and $\nu$ calibrated to ensure that the wind does not start before the maximum in the SFR(t) is reached,
model leoii.sfr.
i.e., we decrease the model SFR with respect to that inferred by HGV, to see how far off we end up.
The value of $\nu$ required in this case is 0.24. This means that the inferred SFR(t) would have to be 
off by a factor of 4, for the tidally limited dark halo of this galaxy to retain the gas long enough
to explain the age spread of stars in Leo II. Since the SFR(t) of HGV was normalized using the total
observed luminosity of these galaxies, such a large error is not possible, the disagreement of model 
leo.dm cannot be explained within the uncertainties of the inferred SFR.

The third model for Leo II, leoii.wind, is calculated by taking $f_{DM}=1.0$ $\nu =1.0$, and
optimizing $\gamma$ so that the wind starts not before the maximum in the SFR is reached. In this way,
we estimate the fraction of energy (and metals) which the SNea's must have lost without interacting
with the ISM of Leo II, all other parameters fixed at their observed values. This yields
 $\gamma=0.82$, which implies that a very significant fraction of the SN ejecta in Leo II must
have left the galaxy without having interacted with the ISM. In view of the models of
Mac Low \& Ferrara (1999) of expanding super bubbles in small galaxies, leading to fountains and 
loss of metals and energy, this large factor seems unlikely. 

Finally, the fourth row gives the results of a model where it was the initial luminous to dark matter
ratio that was changed, again, the high values required for this parameter makes it unlikely that this
be the explanation to the retention of the gas up until the formation of all the visible stars.

Figure 1 shows the time evolution of the model adjusted to give accordance
with the HGV star formation history for Leo II, adjusting only the parameter $f_{DM}$, leoii.dm,
which can be seen in panel (a). The age and duration of the star formation episode closely matches HGV,
until the development of the wind, giving a single episode of star 
formation about 8 Gyr ago, with a duration of around 2 Gyr. A slight re-normalization was hence 
included, as discussed above. Panel (b) shows the evolution of the mean metallicity for the gas (up to
the point when this is lost) and stars (for stars in the $0.5-2 M_{\odot}$ range, those visible today), 
seen to remain constant once star formation ceases. The shaded band shows the data by Mateo (1998)
and Mighell \& Rich (1996), with the central values and error ranges of this determination 
(intrinsically devoid of temporal information, as it is hard to know the ages of the stars whose 
spectra were taken), arbitrarily placed at t=15 Gyr. All observations and predictions throughout 
this work have been consistently scaled to Grevesse \& Sauval (1998) solar abundances -henceforth GS. 

It is clear that our model presents abundant star formation at the
metallicity ranges implied by the observations, although the final present day averages fall slightly
above the observations. In this way, we would identify the stars sampled by the metallicity studies as 
having formed at times during which the model shows corresponding values of the metallicity, coinciding
with the main episode of star formation. Local metallicity variations due to a less than perfect mixing 
of the ejected metals is of course also possible in these galaxies. Other explanations for the higher
average present day metallicities given by our models might be the selective expulsion of metals over
gas. Mac Low \& Ferrara (1999) conclude through careful hydrodynamical simulations of expanding hot
bubbles in dSph systems, that metals produced by the SN explosions are lost much more easily than
the ambient gas of the galaxy. In fact, for the mass range covered by dSph's, these authors find that
never is more than a few percent of the metals produced retained, while under certain conditions up to 
20\% of the gas might remain bound. The models of galactic winds (e.g. Leitherer \& Heckman 1995, 
Dahlem, Weaver \& Hecman 1998, Tenorio-Tagle \& Munzo-Tunon 1998) reviewed in Heckman (2001), also
consistently show not only large winds developing in systems in the dSph range, but also the selective
clearing of metals over ambient gas. A selective ejection of metals similar to what is found in 
hydrodynamical simulations would certainly reconcile our results with observed values. The fractions of 
metals which each model would require to be lost in order for the final average metallicities to match 
observed values, is given by the parameter $\Delta\gamma_Z$, where the range in this parameter
corresponding to the error bars seen in the observational determinations of the metallicities for
our galaxies. Clearly, models .sfr and .wind are already very close to the metallicity constraints,
as very low values of extra metal loss are needed to reach agreement with the upper ranges of the
measured metallicities.

The difference noted above might also be reduced if the presence of metallicity gradients within the
dSph's was included. Tamura, Hirashita \& Takeuchi (2001) model the enrichment of systems like
the local dwarf spheroidals, and conclude that internal metallicity gradients are to be expected,
a difference between the global mean metallicity and local values, corresponding perhaps to the
small regions studied by direct spectroscopic samples, would not be surprising. Still, Shetrone
et al. (2001b) find no evidence of any stars more metal rich than [Fe/H]=-1.45 in the region of 
Ursa Minor which they study. Although this does not prove no stars more metal rich than the above 
limit exist in Ursa Minor, it does make our result of the average present day metallicity for 
this system of [Fe/H]=-1.2 appear unlikely. The most probable explanation being the selective 
expulsion of metals discussed previously, and not directly modeled in our study.

This first test gives us some 
confidence on the validity of our assumptions, as a prediction below the observed range would have 
invalidated the model.

It is intuitive that $(f_{DM}-1) + \gamma  + \nu \approx 1$, if we wanted to calibrate $\gamma$ from the 
currently available metallicities at $\gamma =0.5$, we would therefore end up with $f_{DM}=1.5$, an
interesting possibility. The above makes sense if this value for $\gamma$ is thought of as a
geometric form factor allowing for a disk-like structure for the gas in our systems, which permits 
a fraction of the heated gas to escape before interacting with the totality of the galaxies ISM.

\begin{figure}
\epsfxsize=8.5cm
\epsfbox{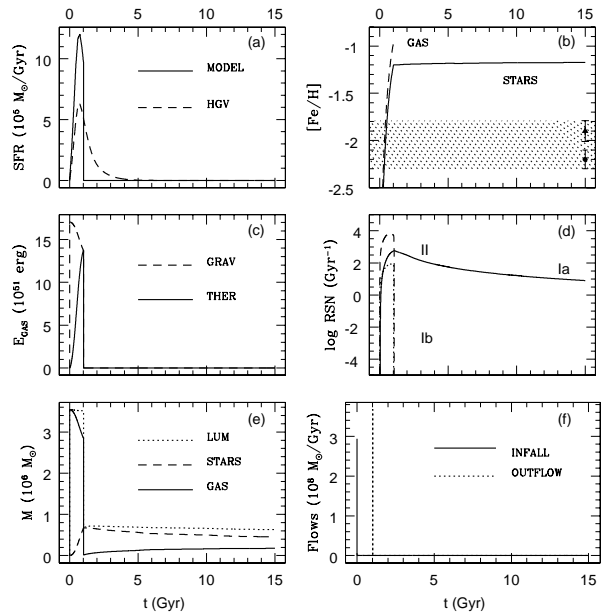} 
{\caption[]
{
Predictions from um.dm model. All panels are totally analogous to those of figure 1.
Data from Mateo (1998) (circle) 
and Shetrone et at. (2001a) (triangle), scaled to solar abundances as by GS.
}
\label{f2}}
\end{figure}

Panel (c) gives
the time evolution of the gravitational and thermal energies of the gas, the first growing due 
to the accretion of gas, and the second due to the heating produced by SNea. The crossing point
of the two curves defines the onset of the galactic wind, and the cessation of star formation.
In panel (d) we show the time evolution of the different SNae rates, with type II and type Ib rates 
ending once star formation has finished, and type Ia rates continuing beyond, reflecting the extended 
evolution of low mass binary systems.
In panel (e) we give the evolution of the gas, stellar and total baryonic masses, with the gas mass
going rapidly to zero with the onset of the galactic wind, and the stellar mass remaining essentially
constant after this point. Finally, panel (f) gives the details of the infall used and outflows 
obtained. The model is largely insensitive to the details of the former, as the latter are a solid 
consequence of having formed as many stars as are visible today, within the inferred dark haloes of 
the observed systems.

\begin{figure}
\epsfxsize=8.5cm
\epsfbox{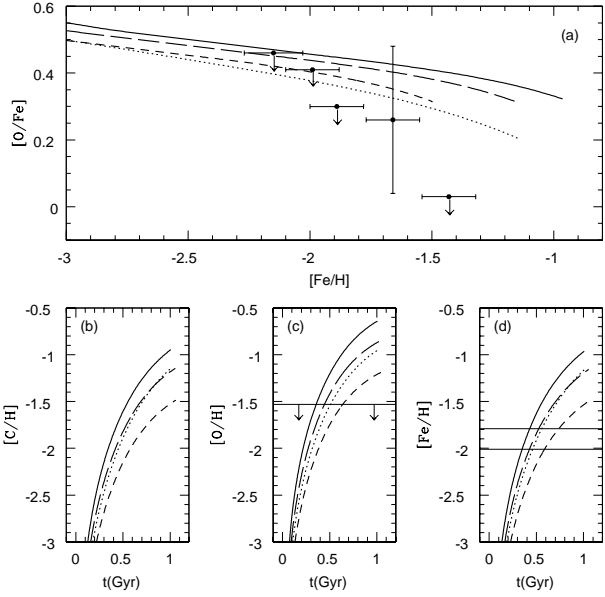} 
{\caption[]{The C, O, and Fe gas abundance evolution of Ursa Minor.
Predictions from um.dm (continuous lines), um.bar (long-dashed lines), um.sfr (short-dashed lines)
and um.wind (dotted lines). Observational data are from Shetrone et al. (2001a). 
Solar abundances as in GS.}
\label{f3}}
\end{figure}

The second set of models in Table 1 gives the corresponding models for Ursa Minor, all analogous to 
those described for Leo II. For this galaxy it is well known that star formation ended 
more than 10 Gyr ago (e.g. Mateo 1998), this is reflected in its $SFR_{HGV}$ which shows a strong 
pulse of star formation in the remote past, and nothing since.
 In this case, the discrepancy in the first model compared to $f_{DM}=1.0$
being 1.32. This yields a higher  $\nu =0.31$ in the second model, a lower
$\gamma=0.74$ in the third and a slightly lower initial luminous to dark matter ratio of 0.24 in 
the fourth model. This numbers remain unacceptably extreme, and again point towards a tidal radius for
this galaxy slightly larger than current estimates.  Figure 2 is totally analogous to Figure 1, and 
gives the details and evolution of model um.dm. Again, we see that the observationally derived band for 
the metallicity of this galaxy coincides with what our model predicts applied during the epoch of most
intense star forming activity. The present day values for the model metallicities being significantly
above the observed range however, perhaps indicative of the reality of this galaxy falling between 
models um.dm and um.wind.

It is interesting that for Ursa Minor we have a further constraint on the models, in the form of the 
details of elemental abundance ratios, from the study of Shertone, Cote \& Sargent (2001). In Figure 3
we show the evolution of [O/Fe] vs. [Fe/H], panel (a), and the detailed temporal evolution of [C/H], 
[O/H] and [Fe/H], panels (b), (c) and (d), respectively. The available observations appear as points 
with error bars, some only upper limits, and error bands. This more delicate record of the relative 
importance of different type of SN events, and of the accretion of fresh material appears to agree 
with the trends obtained from our models, in the [Fe/H]$<$-1.6 range. The time axis in these graphs 
extends only to 1 Gyr after 
the formation of Ursa Minor, since beyond this point no further stars are formed. The solid curves 
are results for model um.dm, with the um.sfr, um.wind and um.bar models corresponding to the 
short-dashed, 
dotted and long-dashed curves, respectively. Notice that the highest  values of the metallicities 
are obtained for model um.dm, since having taken a dark halo that retains all the
gas, it is the model which retains the most metals, and hence results in the highest metallicities. 
For the [Fe/H]$>$-1.6 range, measurements of [O/Fe] are definitively inconsistent with our results,
a problem to which we will return at the end of section 3.

\subsection{Complex Galaxies}

\begin{table}
\caption{Model Parameters for Complex Galaxies}
\begin{center}\scriptsize
\begin{tabular}{lccccc}
Model &  $f_{DM}$ & $\nu$ &  $M_{lum}/M_{DM}$ & $\tau$ & $\Delta\gamma_Z$\\
\hline
\hline
&& Carina &&& \\
carina.dm  & 1.080 & 1.00 & 0.090 & 3.0 & [0.63, 0.86]\\
carina.sfr & 1.000 & 0.63 & 0.090 & 3.0 & [0.60, 0.85]\\
carina.bar & 1.000 & 1.00 & 0.140 & 3.0 & [0.63, 0.86]\\
\hline
&& Leo I &&& \\
leoi.dm  & 1.076 & 1.00 & 0.090 & 1.0 & [0.04, 0.86]\\
leoi.sfr & 1.000 & 0.65 & 0.090 & 1.0 & [0.00, 0.82]\\
leoi.bar & 1.000 & 1.00 & 0.130 & 1.0 & [0.00, 0.85]\\
\hline
\hline
\end{tabular}
\end{center}

\begin{flushleft}
All symbols as in table 1.
\end{flushleft}

\end{table}

\begin{figure}
\epsfxsize=8.5cm
\epsfbox{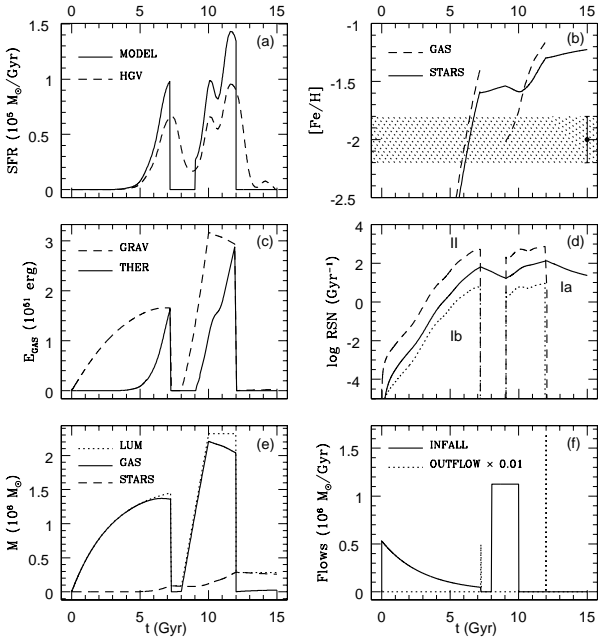} 
{\caption[]{
Predictions from carina.dm model.
Observational data are from Mateo (1998)}
\label{f4}}
\end{figure}

Here we describe our results for Carina and Leo I, galaxies showing a series of
SF episodes. It has been known for some time that these galaxies show evidence of
a complex and extended star formation history, extending across most of the lifetime
of the systems (e.g. Mateo 1998). In HGV, this was confirmed by results showing
two distinct periods of star formation in both galaxies, with a lull in between, each period
having lasted for about 2-3 Gyrs. This behaviour can not be explained within the single 
accretion/violent wind models with which the previous galaxies were described. For these next two
galaxies, we needed to use two episodes of gas accretion, some generalities of the permitted parameter
space are described below. No .wind models were calculated, as their outcome can be inferred from that
of models .dm and .sfr, and from the formula given in the previous section that relates $f_{DM},
\gamma$ and $\nu$.

Table 2 shows in the first set of rows model having as constraints the $SFR_{HGV}$ for the
Carina dwarf. In the first model, the size of the dark halo was adjusted so that the first wind
develops only after the first episode of star formation has ended, much the same as in the previous
two cases. The second star formation episode then determines the characteristics of the second gas
accretion period, the dark halo structure having been fixed by the onset of the first wind. We see that
only very marginal corrections to the observational determinations of the tidal radius of Carina are
required to explain the retention of gas, and take this as evidence in favour of this model. In the 
second row we have model carina.sfr, where the total star formation rate used had to be multiplied 
by 0.63, in order for the inferred dark halo of Carina to retain its gas throughout the long star 
formation activity seen. This large reduction is in conflict with the limits given by the total 
luminosity of the system, and hence we rule out such a model. The third row shows the necessary
adjustment to the $M_{lum}/M_{DM}$ parameter needed for Carina not to develop winds before each 
star formation episode is completed, again, unrealistically high numbers. A carina.wind model would 
give a $\gamma=0.37$, only marginally acceptable from the point of view of more complete models of 
gas heating in small systems, although it would yield present day average metallicities matching the 
observed central values.

\begin{figure}
\epsfxsize=8.5cm
\epsfbox{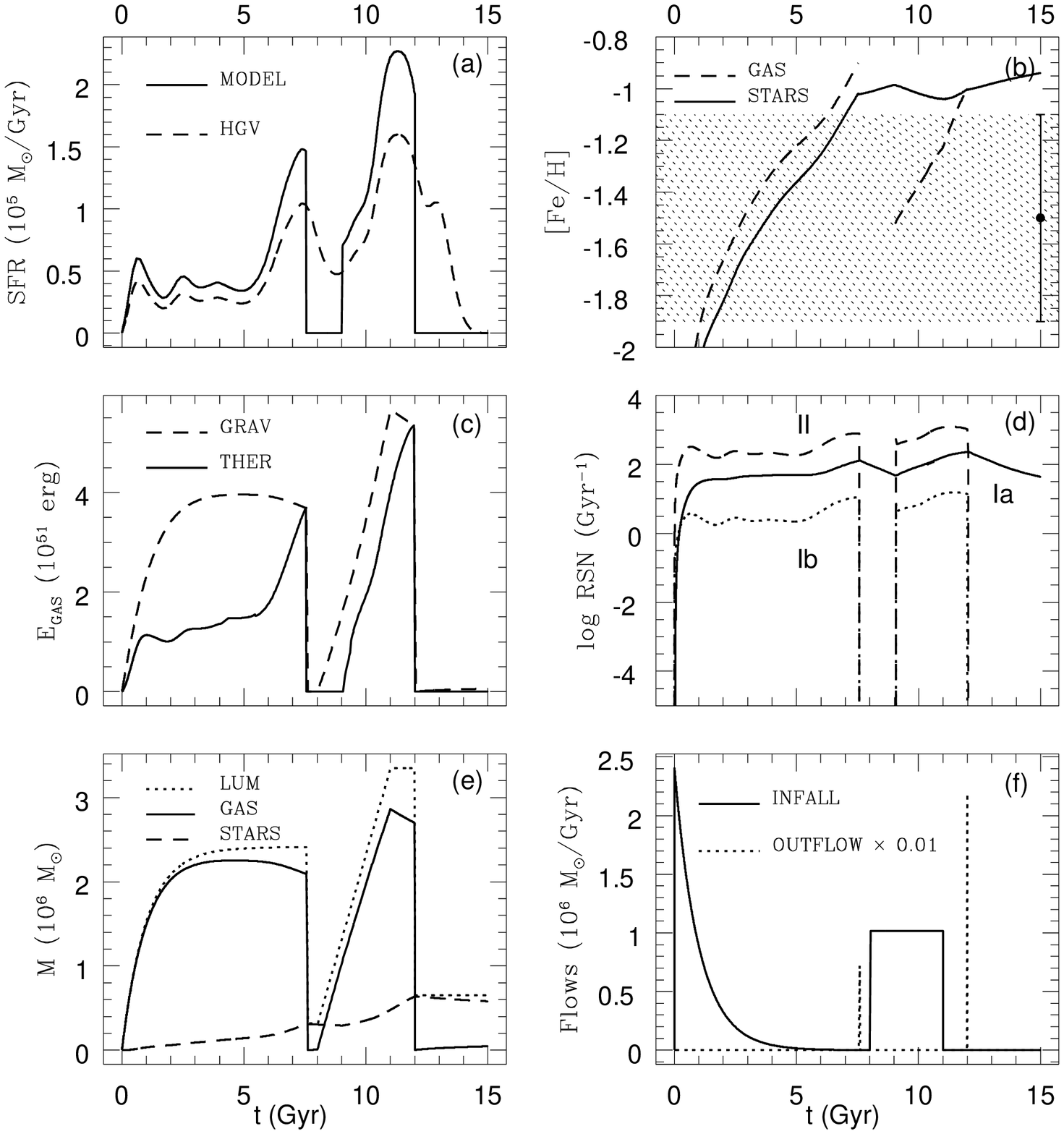} 
{\caption[]{
Predictions from leoi.dm model.
Observational data are from Mateo (1998).}
\label{f5}}
\end{figure}

The temporal evolution of model carina.dm is shown in figure 4, 
with the different panels and curves being
totally analogous to those of figure 1. In panel (a) we see the $SFR_{HGV}$, and the one used by the 
model, having very similar time structures, except for the total cessation in star forming activity
between the two main bursts assumed by the model, 
and not seen in the directly inferred star formation history. This probably 
reflects second order effects, not contemplated by the simple modeling attempted here, which however, 
does include the broad behaviour of the galaxy, and hence we expect to give valuable constraints on the
physics of the evolution, if only at an approximate level. 

In panel (b) we see that the average metallicity of the stars falls a little 
during the second accretion phase, which we take as being composed of primordial material. Again, the
observed metallicity range falls well within the gas metallicity during the two periods of star 
formation, so that our model comfortably accounts for the presence of a significant number of stars 
within the observed range.

Panel (c) shows clearly how the star formation process is limited by the appearance of galactic winds, 
once the thermal energy of the gas surpasses the gravitational potential energy. In Panel (d) we see the
rates of the different types of SNea, with the memory of the first star formation episode affecting the
dynamics of the second, through the extended SNIa rates. Panel (e) gives the cumulative amounts of
stars and gas present, with the gas content being totally cleared off after the second wind. This 
is seen in the final panel, where the two accretion phases and the two galactic winds are shown.

Leo I is very similar to Carina in its star formation history, this is reflected in the second set of
models shown in table 2, where all numbers very closely follow what was obtained for Carina. In figure
5 we show the temporal evolution of model leoi.dm. This figure again closely resembles the results of 
Carina, with the only difference being accordance in the predicted average metallicities and observed
values. Our physical assumptions are shown to be consistent with observational data in predicting the 
total clearing of the gas from this galaxy. Bowen et al. (1997) find no gas around Leo I using three 
lines of sight towards distant QSOs, and searching for absorption features in the spectra.

The values of $f_{DM}$ slightly above unity, i.e. the requirement of a core radius in the dark matter
halo slightly above current observational estimates for the  tidal radii of these systems, is totally
consistent with the very recent dynamical studies of Kleyna et al. (2001) and the photomertic surveys of
Odenkichen et al. (2001) yielding precisely this conclusion.

Here the values of extra metal loss required by the models to agree with the ranges of observed
metallicities, are given by $\Delta\gamma_Z$. We see that for Carina, in all cases, 
we require between 60\%
and 85\% of metal expulsion for our models to agree with the data. In the case of Leo I, very little
of this effect is needed to agree with the upper limits of the measurements, although high values
of 85\% are again needed to reach the lower bound in the observed metallicities.

\subsection{Ursa Minor Again?}

\begin{figure}
\epsfxsize=8.5cm
\epsfbox{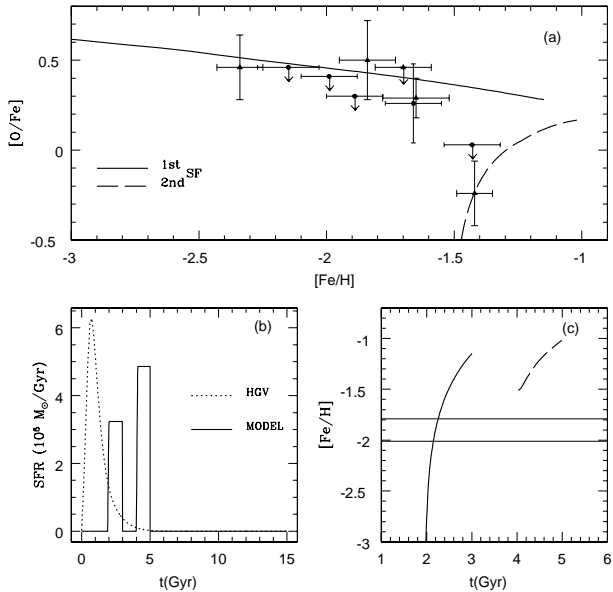} 
{\caption[]
{
Predictions from the Ursa Minor variant model. Panel (a): solid curve gives the [O/Fe] vs. [Fe/H]
ratios for the first burst of the model, with the dashed curve showing results for the second 
(see text).
Data from Shetrone et at. (2001a) (Ursa Minor, circle, Draco, triangle). Panel (b): the star 
formation for the variant Ursa Minor model, showing two distinct bursts, required to match the low 
oxygen abundances measured at high iron abundances for this system. Panel (c): the temporal evolution 
of [Fe/O] in the gas for this model, when SFR$\neq$0.
}
\label{f2}}
\end{figure}

In our treatment of Ursa Minor in section 3.1 we fixed our models in order to comply with $SFG_{HGV}$,
however, as mentioned in the description of our method, the inference of HGV loses all temporal
resolution for ages greater than 10 Gyr. In this way, the time structure inferred for Ursa Minor could
well be only an artifact of the method. As remarked in the description of our present results for
this galaxy, the element ratios we obtain are all similar to what was obtained for Leo II with values
of [O/Fe] always above about 0.2. In comparing with the detailed observational determinations for
Ursa Minor, we find that for values of [Fe/H]$<$-1.6 our corresponding predictions for [O/Fe] closely
match observations for this galaxy. On the other hand, the most metal-rich data point for Ursa Minor of
[Fe/H]$\approx$-1.4 corresponds to an upper limit of [O/Fe] $\approx$ 0.0. 
Our single burst models can not reproduce
such low values of oxygen at such high values of iron. Comparing with results for the ``complex''
galaxies, we found that for those systems, the first wind managed to clear the galaxy of the heavy 
elements produced, which resulted in low values of oxygen in the second burst. The products of the
SNae type II formed during the first burst are cleared away, and hence do not influence the 
metallicities
of stars formed during the second burst. The SNae type Ia due to binary stars formed during the first
burst have such long evolution times though, that they do manage to alter the element ration for stars 
formed during the second burst. In this way, stars formed during the second episode have large amounts
of iron, but little oxygen.
	
This lead us to consider the possibility of a more complex star formation history for Ursa Minor. 
In this sub-section we present an alternative model for Ursa Minor, where we have modeled the star 
formation rate with two discrete bursts, with total amplitudes such that the final luminosity 
matches observed values, and an intensity ratio handled as a free parameter. The duration of the 
two bursts was arbitrarily set to 1 Gyr. Having taken lower absolute intensities for the star 
formation episodes, in this case we obtained $f_{DM}=1.06$, much like in the other ``complex'' 
galaxies, the system can be thought of as self regulated by its dark halo. Only the .dm model 
was calculated in this case.

Figure 6 shows the results for this alternative model for Ursa Minor. The top panel gives the 
observational measurements for the element ratios (dots with error bars and lower limits), 
together with those yielded by the model, shown by the solid line, for the first burst, and dashed one,
for the second one. Also shown are the observations for Draco, triangles with error bars, for 
comparison. Panel (b) shows the star formation rate assumed, compared with that given by HGV, in this 
particular case, indicating only that all stars in Ursa Minor are older than 10 Gyr. Finally, Panel 
(c) shows [Fe/H] for the gas in Ursa Minor during the star formation bursts, according to the 
model. The horizontal lines limit observational results.

This study allows us to raise the interesting possibility that Ursa Minor, the last of the dSph 
galaxies still thought to be characterized by a simple star formation history, a single burst 
long ago, was indeed formed long ago, but quite probably by a more structured star formation history, 
of the type seen in Carina, Leo I and other younger dSph systems. The even more extreme values for 
Draco shown in the element ratios of figure 6 suggest that a multi-event SFR probably also applies 
to this galaxy. 

Although we have taken the low measurements of O at high values of Fe in Ursa Minor to suggest
a complex star formation history for this galaxy, as suggested by models for Carina and LeoI, 
a different possibility also exists. Since O is produced predominantly in type II SNea, and Fe
mostly in SNae type Ia, given that SNea type II typically occur in OB associations, a greater loss
of O over Fe could be expected. In large galaxies, the explosion of a single SN is not sufficient
to expel the material it produced from the galaxy, but the combined effect of several tens of
SNea is clearly enough to expel a significant fraction of the metals produced away from the parent
system. However, this differential expulsion of O over Fe will become increasingly less relevant
as one goes to smaller systems, where the total potential energy becomes progressively smaller and
comparable to the energy input of a single SN event. The effect mentioned above will eventually 
disappear at systems sufficiently small. Given we are treating here the smallest of all galactic
systems, we believe this mechanism for selective loss of O over Fe plays only a marginal role.

This complements our previous results, where the direct
inferences of star formation histories were used to guide and constrain chemical evolution models. In 
this case, it is the chemical evolution model that establishes the guidelines for the star formation
inferences.

\section{Discussion}

We now turn to general aspects of our results, and the interesting connections which in view of them can
be made. Firstly, we note that the final values of the ratio of luminous to dark matter present in all 
the models, lies in the range 0.008-0.022, for the various galaxies and parameter choices. This is the 
range of values inferred for the systems modeled, from direct observations of stellar velocity 
dispersion and total luminosities (Mateo et al. 1993), which gives us confidence that no great errors
have been made in the modeling.

\begin{figure}
\epsfxsize=8.5cm
\epsfbox{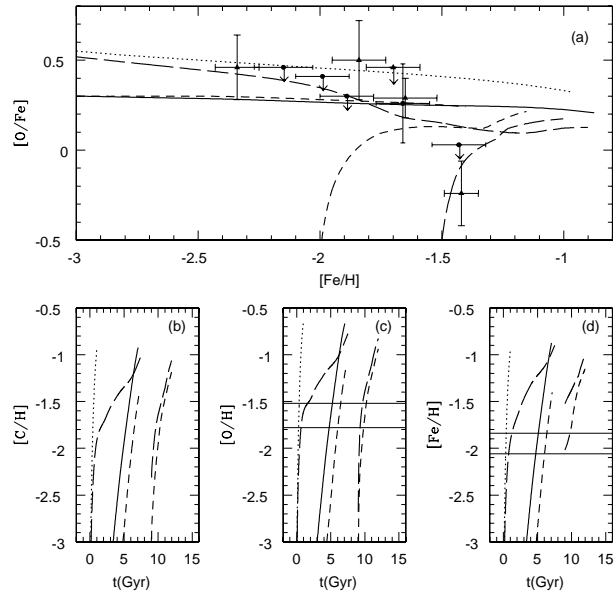} 
{\caption[]{The C, O, and Fe gas abundance evolution for our dSph models.
Predictions from leoii.dm (continuous lines), leoi.dm (long-dashed lines), carina.dm (short-dashed lines)
and um.dm (dotted lines). Observational data are from Shetrone et al. (2001a).
circles: ursa minor, triangles: draco. Solar abundances as in GS.}
\label{f6}}
\end{figure}

As we have a detailed treatment of the stellar yields, we can predict not only total metallicities
as expressed through [Fe/H] ratios, but also details of the different elemental ratios. In figure 7 we 
give the time evolution of [O/Fe], [C/H], [O/H] and [Fe/H], in panels (a), (b), (c) and (d),
respectively. The results for the leoii.dm model shown by the continuous curves, long-dashed for
leoi.dm, short-dashed for carina.dm (the two episodes of star formation giving a discontinuity in the 
curves for this last two) and dotted for the um.dm model. The points with error bars and ranges marked 
are the same as in figure 3, data taken for Ursa Minor, the only one of our galaxies for which such 
studies exist. Data for Draco, a galaxy not studied here were also included (triangles), for which 
Aparicio et al. (2001) find a stellar population consisting mostly of an old component, with an extra
up to 25\% of stars with ages in the 2-3 Gyr range. It is therefore reasonable to group this galaxy
with the
``complex'' systems (Mateo 1998). Although differences might well be expected to appear in this 
parameters as a result of the different epochs and structures of the SFH amongst the galaxies of our
sample, it is interesting that our models fall well within the region defined by observations
of Ursa Minor, a point strengthened by the similar values observed in the younger Draco galaxy. This 
point shows our simple physical model to be broadly consistent with the mechanisms of enrichment and 
accretion that shaped the galaxies studied.

Notice also the two most iron rich points in panel (a), their low oxygen abundances imply these stars 
were formed from material heavily enriched by SN type Ia events. Our models for the ``burst'' galaxies 
all give fairly straight lines in panel (a), always at values of [O/Fe] higher than these last two
points show. It is only in models of systems showing extended episodes of star formation that binary
star SNae contribute significantly to the enrichment of gas which still is forming stars, after the 
first wind has cleared away the oxygen contributed by the SNII events of the first star formation 
event. The curves for the gas metallicity during the second phase of activity in the Carina and 
Leo I models neatly encompass the region defined by these last two points. The extended nature of
star formation in Draco makes this result appear natural, with the results for Ursa Minor having 
suggested the developments presented in section 3.3.

The only disagreement between our models and the data is the appearance of a systematic excess
in the predicted [Fe/H] ratios for our galaxies. As we have already discussed, this is most likely
due to a selective expulsion of metals in general from the systems studied. Here we discuss the 
option of a change in the effective yields being responsible for the mismatch mentioned above.
The possibility of massive SN events ending up as black holes and consequently swallowing up an
important fraction of the heavy elements produced has been discussed in the literature, e.g. Maeder
(1992). However, Carigi (1994) studied this possibility carefully in the context of Solar Neighbourhood
chemical evolution models, and found that local data are incompatible with the idea of any 
significant alteration to the effective yields due to the black hole mechanism. In that work it is
seen that the effects of black holes altering the yields upon [Fe/H] is negligible, as Fe is
produced mostly in binary systems of low total mass. The effects upon [O/Fe] ratios on the other hand,
are quite severe, and in the range of values of [Fe/H] of our models, would lower the [O/Fe] values by
0.9. This would hence not solve our problem with high [Fe/H] values, but would put our [O/Fe] ratios, 
which at present show full accord with observations, much lower, and out of the observed range. The 
applicability of the Carigi (1994) analysis of this effect in the context of Solar vicinity chemical 
evolution
models to our dSph case is guaranteed by the fact that abundance ratios depend mostly upon yields and
IMF, which were essentially the same in both cases, and not upon star formation histories.

Since a very clear distinction appears in the model parameters between the ``burst'' and the ``complex''
galaxies we treated, one can try to trace this clear dichotomy to intuitive physical differences between 
the two sets of objects. A zero order criterion that neatly mirrors this division is given below.

\begin{figure}
\epsfxsize=8.5cm
\epsfbox{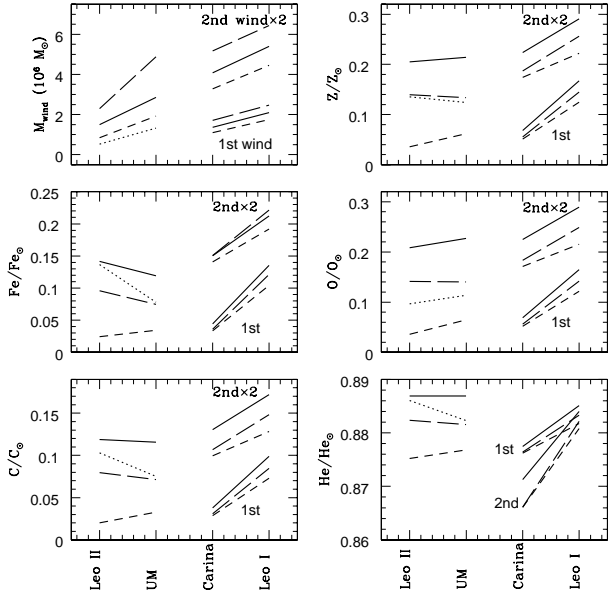} 
{\caption[]{
Ejected Mass through galactic winds and chemical abundances by mass of the expelled material vs dSph.
Predictions from galaxy.dm (continuous lines), galaxy.bar (long-dashed lines), 
galaxy.sfr (short-dashed lines)
and galaxy.wind (dotted lines).
Solar abundances from GS.
}
\label{f6}}
\end{figure}

It is clear that the gravitational potential energy of the gas component will scale with the total 
gravitational energy of the system, which will scale with $M_{total}/R_{total}$. On the other hand,
the thermal energy of the gas will scale with the star formation rate of the system, considering the 
star 
formation episodes of these galaxies to have been of comparable duration, we can expect the criterion
$P=(M_{total}/R_{total})/SFR_{max}$ to divide the sample between the single burst and the complex 
galaxies. Indeed, taking $M_{total}$ equal to the dynamical masses given by Mateo (1998), in units of
$10^{6} M_{\odot}$, $R_{total}$ equal to the inferred tidal radius, in units of kpc, and $SFR_{max}$ 
given by the maximum of the HGV results for each galaxy, in units of $10^{5} M_{\odot}/Gyr$, we obtain
$P$ = 17.0, 16.0, 11.0 and 3.8, for Carina, Leo I, Leo II and Ursa Minor, respectively. Thus, despite 
there being no clear distinction in total masses, luminosities or tidal radii, the parameter $P$ 
clearly 
separates the two groups. We can derive the empirical limit of $P>(12-15)$ as a requirement for a dwarf
spheroidal galaxy to have extended episodes of star formation. 

In this way, knowledge of $M_{total}, 
R_{total}$ and of the integrated luminosity of a dSph galaxy would suffice to give a first indication of
the temporal structure of its SFR, with $SFR_{max}$ estimated from the total luminosity and the age of 
the universe. The division of burst and complex galaxies in our sample through the $P$ criterion, also
suggests that the star forming regime of Carina, Leo I and galaxies of their same class, might have 
actually been regulated mostly by internal processes. Perhaps the secondary accretion episode 
is in fact the re-accretion of the same gas that was ejected by the first wind, plausibly held in an 
extended reservoir. At this point the boundary conditions fixed by the Milky Way would determine 
the details of subsequent accretion events. In the above scenario the metallicities of the second 
wind would be even higher 
than what the model predicts. The inclusion of the Ursa Minor variant model into this criterion 
would imply low values for $SFR_{max}$ in this system, suggesting a very extended SFR history,
all beyond 10 Gyr.

Finally, we present in figure 8 the total ejected masses in the winds of models .dm, for all the 
galaxies. The second winds in Carina and Leo I are shown multiplied by a factor of 2, to distinguish
them from the first ones in the figure. It can be seen that the second wind develops in material which 
has been enriched by the SNIa events formed also during the first burst, together with all SN types of
the second burst. 
We give also the metallicities and elemental ratios of this ejecta, relevant to problems of metal
enrichment in the halo, and in treating the contribution of systems similar to the local dSph's in the
more general context of metal enrichment in clusters and in the high redshift universe, 
were such small systems in fact dominate the 
luminosity function of galaxies. This allows us to identify the ranges of abundances and abundance
ratios likely to be relevant in the problems listed above.

A final caveat must be mentioned in connection with our wind criterion where the energy produced by
the SNae is compared to the total gravitational potential energy of the halo of the systems being
treated. In Mori, Ferrara and Madau (2001) careful hydrodynamical simulations of SN explosions
in small galaxies are performed, reaching the conclusion that only about 30\% of the ejected energy 
is
available to heat and push the ISM around. This result derives mostly from considering radiative
losses, and the effects of the geometry of the problem, SN exploding in the outskirts of the galaxy
will loose a fraction of their energy similar to the solid angle over which the galaxy is not seen.
The precise efficiency factors for SN explosions necessarily depend on the details of how the SN
explosions are distributed, what dark matter profile is adopted and how strongly the SNae are grouped
into associations. The results of Mori, Ferrara and Madau (2001) are however indicative of an 
efficiency
factor of between unity and a third applying to the cases we are treating here. This introduces some
uncertainties into our results, but not enough to change the qualitative picture. For example,
the factors by which the star formation has to be reduced in the burst galaxies would change from
around 0.8 to 0.25, i.e., we would still identify a strong wind developing in these systems, 
consistent with the evidence in favour of winds having cleared the local dSphs of their gas. 
Basically, in the context of the recent hydrodynamical simulations mentioned above, the masses of the
ejected winds of our figure 8 are probably uncertain and have to be multiplied by a factor of 
between 1.0 and 0.3.

\section{Conclusions}

To conclude:

1) We have performed a detailed exploration of the parameter space available to the four 
local dSph galaxies studied here, in terms of the gas accretion regimes and metallicities of the 
formed stars and ejected materials,
by taking as external restrictions the star formation histories of these systems, as inferred from direct
statistical studies of their resolved populations, together with the observed general properties of their
dark haloes. This shows that knowledge of the time structure and normalization of the star formation
rate of external galaxies can be combined with physical and chemical modeling of these systems to derive
interesting information on their evolution and dark matter haloes, the details of which furnish the
boundary conditions within which the galaxies evolve.

2) From the observed abundance ratios of Ursa Minor, in combination with the physics of gas flows
and our chemical models, we find strong suggestions that this galaxy, often thought of as the 
prototypical old, single burst and low metallicity dSph, might in fact have experienced a complex
star formation history. This has remained hidden from direct studies of its resolved stellar 
population by its large age, i.e. everything happened more than 10 Gyrs ago, but it was not simple.

3) We found evidence for a slight underestimate of $R_{tidal}$ as the total extent of their dark halos
or of a significant metal rich ejecta, with reality probably falling somewhere in between.

4) Comparison of the predicted abundance ratios with the available data shows a broad consistency of our
chemical and physical modeling with the relevant observations.

5) A simple physical 
criterion to estimate when a dSph system might sustain extended star formation, as opposed to being 
subject to a single burst of activity is presented, which neatly separates into those two classes the 
galaxies we studied.

\section*{\bf ACKNOWLEDGMENTS}
The authors wish to thank the referee, Andrea Ferrara, for a very careful reading of the manuscript, 
and for helpful suggestions which improved the clarity and quality of the presentation.

The work of L. Carigi was partially supported by DGAPA/UNAM through project IN-109696. 
Carigi thanks the Institute of Astronomy, Cambridge, for a summer visitors grant.

\end{document}